\documentclass[english]{article}
\usepackage[T1]{fontenc}
\usepackage[latin9]{inputenc}
\usepackage{amsmath}

\makeatletter
\usepackage{amssymb}
\usepackage{latexsym}
\usepackage{epsfig}
\usepackage{float}
\usepackage{dsfont}

\newcommand{\be}{\begin{equation}}
\newcommand{\ee}{\end{equation}}

\allowdisplaybreaks

\makeatother

\usepackage{babel}
\begin{document}
{}~ \hfill\vbox{\hbox{CTP-SCU/2017007}}\break
\vskip 3.0cm
\centerline{\Large \bf Local geometry from  entanglement entropy}
\vspace*{10.0ex}
\centerline{\large Peng Wang$^a$, Houwen Wu$^{a,b}$ and Haitang Yang$^a$}
\vspace*{7.0ex}
\vspace*{4.0ex}
\centerline{\large \it $^a$Center for theoretical physics}
\centerline{\large \it Sichuan University}
\centerline{\large \it Chengdu, 610064, China} \vspace*{1.0ex}
\vspace*{4.0ex}
\centerline{\large \it $^b$Center for the Fundamental Laws of Nature}
\centerline{\large \it Harvard University}
\centerline{\large \it Cambridge, MA 02138 USA} \vspace*{1.0ex}

\centerline{pengw@scu.edu.cn, wu\_houwen@g.harvard.edu, hyanga@scu.edu.cn}

\vspace*{10.0ex}
\centerline{\bf Abstract} \bigskip \smallskip
Constructing the corresponding geometries from  given entanglement entropies of a boundary QFT is a big challenge and leads to the grand project \emph{ it from Qubit}. Based on the observation that the AdS metric in the Riemann Normal Coordinates (RNC) can be summed into a closed form, we find that the AdS$_3$  metric in RNC can be straightforwardly read off from the entanglement entropy of CFT$_2$. We use the finite length or finite temperature CFT$_2$  as examples to demonstrate the identification.

\vfill 
\eject
\baselineskip=16pt
\vspace*{10.0ex}

Quantum entanglement is one of the most distinct features of quantum
systems. When a quantum system is divided into two parts, the natural
way to measure the correlation between these two subsystems is to
calculate the Entanglement Entropy (EE). We divide the system under
consideration into two regions: $A$ and $B$. The total Hilbert space
is therefore decomposed into $\mathcal{H}=\mathcal{H}_{A}\otimes\mathcal{H}_{B}$.
Then, one traces out the degrees of freedom of region $B$ to get
the reduced density matrix of region $A$: $\rho_{A}=\mathrm{Tr}_{\mathcal{H}_{B}}\rho$.
The entanglement entropy of the region $A$ is evaluated by the von
Neumann entropy, $S_{A}=-\mathrm{Tr}\left(\rho_{A}\ln\rho_{A}\right)$.
It is widely believed that the entanglement entropy is of great help
to realize AdS/CFT correspondence \cite{Maldacena:1997re}, the correspondence
between the quantum gravity in the bulk of AdS and gauge theory on
the AdS conformal boundary. Though no proof for this correspondence
exists so far, many evidences have been proposed to support this conjecture.
Based on AdS/CFT, a holographic way to calculate the entanglement
entropy from the dual gravity theory was proposed by Ryu and Takayanagi
(RT) \cite{Ryu:2006bv} and has been verified extensively on AdS$_{3}$/CFT$_{2}$,
referring to a recent review \cite{Rangamani:2016dms} and references
therein. To approach the entanglement entropy from the dual gravity
side, RT generalized the well-known Bekenstein-Hawking entropy $S_{BH}$
of black hole \cite{Bekenstein:1973ur,Hawking:1974sw} :

\begin{equation}
S_{BH}=\frac{\mathrm{Area\left(\mathrm{horizon}\right)}}{4G_{N}^{\left(D\right)}},
\end{equation}

\noindent which turns out to be the same as the entanglement entropy
$S_{A}$ of region $A$ on the boundary where the conformal theory
lives. To be specific, considering an AdS$_{d+1}$ , on the boundary
$\mathcal{B}$, taking a spatial slice $\Sigma_{\mathcal{B}}\subset\mathcal{B}$,
for a subset $A\subset\Sigma_{\mathcal{B}}$, whose boundary is $\partial A$,
a minimal surface $\gamma_{A}$ in the bulk of AdS$_{d+1}$ can be
identified. The minimal surface $\gamma_{A}$ shares the same boundary
of $A$ , namely $\partial A=\partial\gamma_{A}$. Then, comparing
with the Bekenstein-Hawking entropy, one can treat the area of the
minimal surface $\gamma_{A}$ as a ``horizon'' of $A$. Thus we
can define the entanglement entropy of $A$ as

\begin{equation}
S_{A}=\frac{\mathrm{Area\left(\gamma_{A}\right)}}{4G_{N}^{\left(D\right)}}.\label{eq:EE}
\end{equation}

Since the entanglement entropies calculated from the holographic CFT
and dual gravity are in agreement as expected from AdS/CFT, it is
natural to ask if we can construct the dual geometry from given entanglement
entropies. This idea leads to the grand project \emph{it from Qubit}.
However, when we calculate the area of minimal surface in the bulk,
the geometric structure, specifically the metric, is lost. It is then
difficult to realize the relation between the spacetime geometry and
quantum entanglement. In \cite{Swingle:2009bg,VanRaamsdonk:2009ar,VanRaamsdonk:2010pw},
the authors made efforts to build connections between the spacetime
geometry and the quantum entanglement in QFT. This idea was further
developed by Maldacena and Susskind \cite{Maldacena:2013xja}. They
conjectured an equivalence between Einstein-Rosen bridge (ER) and
Einstein-Podolsky-Rosen (EPR) paradox. In a recent work \cite{Czech:2015qta},
the authors made a nice progress on this direction by introducing
the concept of kinematic space, defined on the oriented geodesics
of AdS$_{3}$. After identifying the Crofton form with the second
derivatives of the given entanglement entropy of CFT$_{2}$, they
read the metric of dS$_{2}$ which is the kinematic space of AdS$_{3}$.

Based on our previous work \cite{Wang:2017mpm}, in this paper, without
any auxiliary fields, we give a transparent construction of the local
geometry from a given entanglement entropy, specifically, AdS$_{3}$
metric as an instance. To start with, let us recall the strategy of
RT proposal in AdS$_{3}$/CFT$_{2}$. To calculate the holographic
entanglement entropy, on the geometry side, one needs to calculate
the area of the corresponding minimal surface. In AdS$_{3}$ static configuration, the
minimal surface is simplified as geodesics. Therefore, geodesics play
a central role in this formalism. On the other hand, it is inspiring
to notice that geodesics are the basis of the Riemann Normal Coordinates
(RNC). We are thus led to conjecture RNC may be of use to construct
the geometry from given entanglement entropy. Moreover, the holographic
entanglement entropy is usually determined by two length scales, $l$
of the visible subsystem $A$ and $L$ of the whole system under consideration.
While in RNC, the metric for a point $P$ is determined by its distance
from the origin $O$ and the scale of the whole bulk geometry.

Remarkably, it turns out indeed we can read the RNC metric straightforwardly
from the entanglement entropy. This is possible by the observation
that in RNC, the perturbatively expanded metric can be summed into
a closed form for maximally symmetric spaces. Moreover, the sum for
AdS$_{d+1}$ matches precisely the string worldsheet topology sum
in flat spacetime. We are thus led to conclude in \cite{Wang:2017mpm}
that AdS genus zero worldsheet corresponds to all genus expansion
of string theory in flat spacetime. This correspondence connects geometry
and topology. 

Let us demonstrate how this works in RNC. On a manifold $M$, taking
a point $O\in M$, any vector in the tangent space $V\in T_{O}M$
defines a unique geodesic $\gamma(s)$ passing through $O$ by the
exponential map exp$:T_{O}M\to M$. The tangent vector of the geodesic
at $O$ is $V$,

\[
\gamma(0)=O,\qquad\frac{d\gamma}{ds}|_{s=0}=V,\qquad\exp(V)=\gamma(1).
\]
Then set the coordinates $x^{\mu}(s)=sV^{\mu}$ to build the RNC.
From the construction, the RNC takes geodesics as coordinates and
we have $g_{ij}(O)=\delta_{ij}$, $\partial g_{ij}(O)=0$. It is known
that RNC fails around singularities where geodesics cannot reach (geodesically
incomplete), or the neighborhood of some point where different geodesics
emanating from the origin $O$ cross. So at least for manifolds we
are concerned like AdS, the RNC is well defined for finite regions
in the bulk. Without loss of generality, we set the coordinate of
the origin $O$ as $0$ and expand the metric in the defining region,
which is not necessarily local. The Taylor expansion is greatly simplified
in RNC

\begin{eqnarray}
g_{ij}^{RNC}\left(X\right) & = & \delta_{ij}+\frac{1}{3}R_{iklj}X^{k}X^{l}+\frac{1}{6}D_{k}R_{ilmj}X^{k}X^{l}X^{m}\nonumber \\
 &  & +\frac{1}{20}\left(D_{k}D_{l}R_{imnj}+\frac{8}{9}R_{iklp}R_{\;mnj}^{p}\right)X^{k}X^{l}X^{m}X^{n}+\ldots.\label{eq:metric expansion}
\end{eqnarray}

\noindent We now  set the background as a $d+1$-dimensional 
AdS spacetime which is a maximally symmetric space with $D_{m}R_{ikjl}=0$
and $R_{ikjl}=-\frac{1}{R_{AdS}^{2}}\left(g_{ij}g_{kl}-g_{il}g_{kj}\right)$. We choose Euclidean signature  because  what we are going to compare with is the static case of the CFT$_2$ entanglement entropy.  
It is remarkable that, referring to the derivations in the Appendix of \cite{Wang:2017mpm},
the expansion can be summed over as a closed form,

\begin{eqnarray}
g_{ij}^{RNC}(X) & = & \delta_{ij}+\frac{1}{2}\sum_{n=1}^{\infty}\frac{2^{2n+2}}{(2n+2)!}\delta_{im}\ell^{m}\,_{a_{1}}\ell^{a_{1}}\,_{a_{2}}\cdots\ell^{a_{n-1}}\,_{j}\nonumber \\
 & = & \left[\frac{\sin^{2}\left(\frac{\ell}{R_{AdS}}\right)}{\left(\frac{\ell}{R_{AdS}}\right)^{2}}\right]^{a}\,_{i}\,\delta_{aj},\label{eq:RNC metric}
\end{eqnarray}

\noindent where we defined

\begin{equation}
\left(\ell^{2}\right)^{a}\,_{b}\equiv-\delta_{b}^{a}X^{2}+X^{a}X_{b}.\label{eq:para define}
\end{equation}

\noindent Or we can write the metric in a more explicit but less compact
form 

\begin{equation}
g_{ij}^{RNC}(X)=\left(\delta_{ij}-\frac{X_{i}X_{j}}{X^{2}}\right)\frac{\sin^{2}\left(\frac{|X|}{R_{AdS}}\right)}{X^{2}/R_{AdS}^{2}}+\frac{X_{i}X_{j}}{X^{2}},\label{eq:RNC metric-1}
\end{equation}
where the indices are raised and lowered by $\delta_{ij}$ and $|X|=\sqrt{X^{2}}$.
It is very interesting to note that the distance $|X|$ is not measured
by the AdS geometry, but by the flat metric, which again motives us
to identify it with the length scale of the subsystem on the boundary
in the following discussions.

We now look at the entanglement entropies calculated from CFT$_{2}$.
It is not easy to compute the von Neumann entropy directly in CFT.
Instead, the replica trick is employed to calculate Renyi entropy
$S_{A}^{\left(n\right)}$ \cite{Caraglio:2008pk,Calabrese:2004eu}
and the entanglement entropy is obtained as follows

\begin{equation}
S_{EE}\equiv\underset{n\rightarrow1}{\lim}S_{A}^{\left(n\right)}=\underset{n\rightarrow1}{\lim}\left[\frac{1}{1-n}\ln\left(\mathrm{tr}\rho_{A}^{n}\right)\right].
\end{equation}

In this work, we are focused on the static cases. The vacuum entanglement
entropy of CFT in a finite region, corresponding to pure AdS$_{3}$,
is given by 

\begin{equation}
S_{EE}^{vac}=\frac{c}{6}\ln\left[\frac{\sin^{2}\left(\frac{\pi l}{L}\right)}{\left(\frac{\pi l}{L}\right)^{2}}\right]+\textnormal{divergent terms},\label{eq:finite size EE}
\end{equation}
where $l$ is the length of the subsystem under consideration and
$L$ is the length of the total system. It is immediate to see that
one can read the RNC metric (\ref{eq:RNC metric}) directly from this
entanglement entropy under the identifications

\begin{equation}
L\leftrightarrow\pi R_{AdS},\qquad l^{2}\leftrightarrow\ell^{2}.\label{eq:identification}
\end{equation}

\noindent These two identifications are very physical. The identification
$L\leftrightarrow\pi R_{AdS}$ is inspiring since, in a rough sense,
$L$ represents the circumference of radius $R_{AdS}$ geometry. One
may be puzzled by the identification of $l^{2}$ and $(\ell^{2})^{a}\,_{b}$
since $l$ is a number while the latter is a matrix. This can be easily
understood by looking at eqn. (\ref{eq:para define}) and (\ref{eq:RNC metric-1}).
Basically, $l$ corresponds to the length $|X|$ of a set of vectors
$X^{i}$'s, which implies that for a specific $l$, we can determine
the metric for all the points with the same distance $l$ from the
origin by using (\ref{eq:para define}). This actually provides an explanation to the
\emph{overdetermined puzzle,} namely, in an asymptotically AdS$_{d+1}$
spacetime, the metric usually is specified by $(d+1)(d+2)/2$ functions,
in contrast to that the bipartition of a Cauchy slice is specified
by two functions. 

In eqn. (\ref{eq:finite size EE}), the divergent term takes a form
of $\frac{c}{3}\ln\frac{l}{a}$, where $a$ is a cut-off. This terms
accounts for the failure of the RNC near the AdS boundary. Therefore,
we even get more information than naively expected, i.e. the existence
of the asymptotic region. 

We turn to another closely related configuration. The entanglement
entropy with finite temperature is given by

\begin{equation}
S_{EE}^{fin}=\frac{c}{6}\ln\left[\frac{\sinh^{2}\left(\frac{\pi l}{\beta}\right)}{\left(\frac{\pi l}{\beta}\right)^{2}}\right],
\end{equation}

\noindent where $\beta=\frac{1}{T}$ is an inverse of temperature and we dropped off the divergent term for reasons explained. It is straightforward to get the corresponding RNC metric by identifying 

\begin{equation}
i\beta\leftrightarrow\pi R_{AdS},\qquad l^{2}\leftrightarrow\ell^{2}.\label{eq:Tidentification}
\end{equation}

\noindent The corresponding global geometry can be easily found 

\begin{equation}
ds^{2}=R_{AdS}\left(-\cosh^{2}\left(\rho\right)dt_{E}^{2}+d\rho^{2}+\sinh^{2}\left(\rho\right)d\theta^{2}\right),
\end{equation}

\noindent Note in order to compare with the static configuration consistently, we are using Euclidean AdS from the very beginning, so there is a minus sign for the $dt_E^2$ term, which is different from the usual conventions in literature. Therefore, this global geometry is nothing but the  Euclidean BTZ black hole. More explicitly, based on the equivalence between the Euclidean BTZ black hole at temperature $T$ and Euclidean AdS$_3$ at temperature $1/T^4$, utilizing the transformation 

\begin{equation}
r=r_{+}\cosh\rho,\qquad\tau=\frac{R_{AdS}}{r_{+}}\theta,\qquad\varphi=-i\,\frac{R_{AdS}}{r_{+}}t_{E},
\end{equation}

\noindent we get the three dimensional Euclidean BTZ black hole

\begin{equation}
ds^{2}=\left(r^{2}-r_{+}^{2}\right)d\tau^{2}+\frac{R_{AdS}}{r^{2}-r_{+}^{2}}dr^{2}+r^{2}d\varphi^{2},
\end{equation}

\noindent where $0<\varphi<2\pi$ and $\tau$ is compactified as $\tau\sim\tau+\frac{2\pi R}{r_{+}}$. 

The  eqn. (\ref{eq:identification}) and (\ref{eq:Tidentification}) show  transparent identifications of the local RNC metric from the dual CFT entanglement entropy. This result is quite surprising. How to interpret it? Let us think about the geometry side. As a \emph{local} theory, basically, all the information about classical gravity is contained in the metric (with Levi-Civita connection). For most  asymptotic AdS geometries we are concerned, the RNC    can be defined in finite region - named as quasi-local. We already see that the RNC metric at point $P$ depends on the location of the origin $O$, in some sense, these two points are ``entangled''. On the other hand, since the metric at $O$ is flat, in other words, the information about the geometry at $O$ is not reachable by solely referring to itself, but is stored in points like $P$. Therefore, in a loose sense,   RNC is the natural language for the \emph{ gravity version of entanglement entropy}.

We take the finite region or finite temperature CFT$_2$ as examples in this work. Since the dual geometry is very simple, we can easily figure out the  global geometries, AdS$_3$ or BTZ black hole respectively. It is very tempting to ask: does this identification between RNC metric and entanglement entropy  also work for other more non-trivial asymptotic AdS geometries? When the holographic entanglement entropy does not take a closed form, the identification of local metric seems still possible. But for the global metric, it may not be straightforward.  It is inspiring to notice that in the RNC metric expansion (\ref{eq:metric expansion}) for any geometry, what really control the expansion are the connections \cite{Wang:2017mpm}. Therefore, it looks possible to construct the global geometry via the connections. The system with a finite size  at finite temperature   for a free Dirac fermion   in two dimensions \cite{Azeyanagi:2007bj} is a good areana to test this conjecture. 

Moreover, based on the fact that $g_{ij}^{RNC}$ must respect the Einstein
equation, it would not be very hard to  construct the dynamical equation for the entanglement entropy through the identification.

The Renyi entropy, as a  trick to calculate the entanglement entropy, has an obscured physical picture. It is calculated from the two point functions of twist fields. We know that the two point functions of CFT$_2$ is completely determined by the distance and the conformal dimension of the operators. From our derivations, the information about the distance on the CFT$_2$ is completely included in the RNC metric. This may provide some clues to build a correspondence between the CFT two point function and some gravity quantities.

In a previous paper \cite{Wang:2017mpm}, we demonstrated a correspondence between genus zero string worldsheet in AdS and the sum of all genus string worldsheet in flat geometry. Comparing with the observation in this paper,  we are led to conjecture there might be another method to calculate the entanglement entropy in CFT, specifically, alike to the Gopakumar-Vafa (GV) \cite{Gopakumar:1998ii,Gopakumar:1998jq}  method, lifting  the theory  to the equivalent M-theory and calculate the entanglement entropy in a physically transparent way. From what we learned from GV, we tend to believe it would represent a sum of the genus expansion of string worldsheets.

\vspace{5mm}

\noindent {\bf Acknowledgements} 
We are especially indebt to Bo Ning for many illuminating discussions and clarifications on the subject. We are also grateful to F. Liu, S. Kim, D. Polyakov and Y. Zhou for  helpful discussions. This work is supported in part by the NSFC (Grant No. 11175039 and 11375121 ). 


\begin{thebibliography}{99}

\bibitem{Maldacena:1997re}    
J.~M.~Maldacena,   ``The Large N limit of superconformal field theories and supergravity,''   Int.\ J.\ Theor.\ Phys.\  {\bf 38}, 1113 (1999)   [Adv.\ Theor.\ Math.\ Phys.\  {\bf 2}, 231 (1998)]   doi:10.1023/A:1026654312961   [hep-th/9711200].   

\bibitem{Ryu:2006bv}    
S.~Ryu and T.~Takayanagi,   ``Holographic derivation of entanglement entropy from AdS/CFT,''   Phys.\ Rev.\ Lett.\  {\bf 96}, 181602 (2006)   doi:10.1103/PhysRevLett.96.181602   [hep-th/0603001].   

\bibitem{Rangamani:2016dms}    M.~Rangamani and T.~Takayanagi,   ``Holographic Entanglement Entropy,''   arXiv:1609.01287 [hep-th].   

\bibitem{Bekenstein:1973ur}    J.~D.~Bekenstein,   ``Black holes and entropy,''   Phys.\ Rev.\ D {\bf 7}, 2333 (1973).   doi:10.1103/PhysRevD.7.2333   

\bibitem{Hawking:1974sw}    S.~W.~Hawking,   ``Particle Creation by Black Holes,''   Commun.\ Math.\ Phys.\  {\bf 43}, 199 (1975)   Erratum: [Commun.\ Math.\ Phys.\  {\bf 46}, 206 (1976)].   doi:10.1007/BF02345020   


\bibitem{Swingle:2009bg}    B.~Swingle,   ``Entanglement Renormalization and Holography,''   Phys.\ Rev.\ D {\bf 86}, 065007 (2012)   doi:10.1103/PhysRevD.86.065007   [arXiv:0905.1317 [cond-mat.str-el]].   

\bibitem{VanRaamsdonk:2009ar}    M.~Van Raamsdonk,   ``Comments on quantum gravity and entanglement,''   arXiv:0907.2939 [hep-th].   

\bibitem{VanRaamsdonk:2010pw}    M.~Van Raamsdonk,   ``Building up spacetime with quantum entanglement,''   Gen.\ Rel.\ Grav.\  {\bf 42}, 2323 (2010)   [Int.\ J.\ Mod.\ Phys.\ D {\bf 19}, 2429 (2010)]   doi:10.1007/s10714-010-1034-0, 10.1142/S0218271810018529   [arXiv:1005.3035 [hep-th]].   


\bibitem{Maldacena:2013xja}    J.~Maldacena and L.~Susskind,   ``Cool horizons for entangled black holes,''   Fortsch.\ Phys.\  {\bf 61}, 781 (2013)   doi:10.1002/prop.201300020   [arXiv:1306.0533 [hep-th]].   


\bibitem{Czech:2015qta}    
B.~Czech, L.~Lamprou, S.~McCandlish and J.~Sully,   ``Integral Geometry and Holography,''   JHEP {\bf 1510}, 175 (2015)   doi:10.1007/JHEP10(2015)175   [arXiv:1505.05515 [hep-th]].   





\bibitem{Wang:2017mpm}    P.~Wang, H.~Wu and H.~Yang,   ``Correspondence between genus expansion and $\alpha^{\prime}$ expansion in string theory,''   arXiv:1703.05217 [hep-th].   

\bibitem{Cadoni:2010kla}    M.~Cadoni and M.~Melis,   ``Entanglement Entropy of AdS Black Holes,''   Entropy {\bf 12}, no. 11, 2244 (2010).   doi:10.3390/e12112244   

\bibitem{Casini:2011kv}    H.~Casini, M.~Huerta and R.~C.~Myers,   ``Towards a derivation of holographic entanglement entropy,''   JHEP {\bf 1105}, 036 (2011)   doi:10.1007/JHEP05(2011)036   [arXiv:1102.0440 [hep-th]].   



\bibitem{Caraglio:2008pk}    
M.~Caraglio and F.~Gliozzi,   ``Entanglement Entropy and Twist Fields,''   JHEP {\bf 0811}, 076 (2008)   doi:10.1088/1126-6708/2008/11/076   [arXiv:0808.4094 [hep-th]].   

\bibitem{Calabrese:2004eu}    P.~Calabrese and J.~L.~Cardy,   ``Entanglement entropy and quantum field theory,''   J.\ Stat.\ Mech.\  {\bf 0406}, P06002 (2004)   doi:10.1088/1742-5468/2004/06/P06002   [hep-th/0405152].   

\bibitem{Azeyanagi:2007bj} 
  T.~Azeyanagi, T.~Nishioka and T.~Takayanagi,``Near Extremal Black Hole Entropy as Entanglement Entropy via AdS(2)/CFT(1),''   Phys.\ Rev.\ D {\bf 77}, 064005 (2008)  doi:10.1103/PhysRevD.77.064005  [arXiv:0710.2956 [hep-th]].

\bibitem{Gopakumar:1998ii}
R.~Gopakumar and C.~Vafa,   ``M theory and topological strings. 1.,''   hep-th/9809187.   

\bibitem{Gopakumar:1998jq}
R.~Gopakumar and C.~Vafa,   ``M theory and topological strings. 2.,''   hep-th/9812127.   




\end{thebibliography}
\end{document}